\documentclass[aps,prl,preprint,superscriptaddress]{revtex4}
\usepackage{graphicx}

\begin{document}
\title{Low-Frequency Microwave Induced Quantum Oscillations in A Two-Dimensional Electron System}
\author{Jian Mi}
\affiliation{International Center for Quantum Materials, Peking University, Beijing, 100871, China}

\author{Huiying Liu}
\affiliation{International Center for Quantum Materials, Peking University, Beijing, 100871, China}

\author{Junren Shi}
\affiliation{International Center for Quantum Materials, Peking University, Beijing, 100871, China}
\affiliation{Collaborative Innovation Center of Quantum Matter, Beijing, 100871, China}

\author{L. N. Pfeiffer}
\affiliation{Department of
Electrical Engineering, Princeton University, Princeton, New Jersey
08544, USA}

\author{K. W. West}
\affiliation{Department of
Electrical Engineering, Princeton University, Princeton, New Jersey
08544, USA}

\author{K. W. Baldwin}
\affiliation{Department of
Electrical Engineering, Princeton University, Princeton, New Jersey
08544, USA}

\author{Chi Zhang}
\altaffiliation{Electronic address: gwlzhangchi@pku.edu.cn}
\affiliation{International Center for Quantum Materials, Peking University, Beijing, 100871, China}
\affiliation{Collaborative Innovation Center of Quantum Matter, Beijing, 100871, China}

%\date{\today}

\pacs{73.43.-f}

\begin{abstract}
  We study the magnetoresistance of an ultrahigh mobility GaAs/AlGaAs two-dimensional electron sample in a weak magnetic field under low-frequency ($f < 20$ GHz) microwave (MW) irradiation.
  We observe that with decreasing MW frequency, microwave induced resistance oscillations (MIRO) damp and multi-photon processes become dominant.
  At very low MW frequency ($f < 4$ GHz), MIRO disappears gradually and a new SdH-like oscillation develops.
  The analysis indicates that the new oscillation may originate from alternating Hall-field induced resistance oscillations (ac-HIRO), or can be viewed as a multi-photon process of MIRO in low MW frequency limit.
  Our findings bridge the non-equilibrium states of MIRO and HIRO, which can be brought into a frame of quantum tunneling junction model.
\end{abstract}

\maketitle
High quality GaAs/AlGaAs two-dimensional electron system (2DES) has been the optimal experimental platform for the study of fractional quantum Hall effect (FQHE) for decades.
Much attention is paid to non-equilibrium transport properties of 2DES in very high Landau levels, such as microwave induced resistance oscillations (MIRO) ~\cite{Zudov2001, YeAPL2001} and zero resistance state (ZRS) ~\cite{Mani, Zudov}, which can be expressed as a function of $\omega / \omega_{c}$, where $ \omega = 2\pi f$ is the microwave (MW) frequency, $\omega_{c} = eB/m^{*}$ is the cyclotron frequency, and $m^{*}$ is the effective mass of electron.
Another notable effect in 2DES is Hall field induced resistance oscillations (HIRO) or Zener tunneling that emerges in a dc-electric field ~\cite{YCL2002, ZudovPRB2007}, and can be described as a function of $2R_{c}/ \Delta Y$, where $2R_{c}$ is the cyclotron diameter, $\Delta Y = \hbar \omega_{c}/eE$ is the real space change due to the inter-Landau level spacing, and $E$ is the Hall electric field.

Experimentally, MIRO and HIRO exhibit very similar features, but their physical origins are different.
MIRO is understood in terms of MW induced impurity scattering (displacement model) ~\cite{Durst2003} and change of distribution function (inelastic model) ~\cite{Dimitriev2005}, while HIRO can be explained by inter-Landau level elastic disorder, which relies on a large momentum transfer ~\cite{YCL2002}.
Many experimental and theoretical efforts are made to research the interaction between MIRO and HIRO ~\cite{HatkePRB2008, ZudovPRL2007, HatkePRL2008}.
It is found that the two oscillations mix nonlinearly when a 2DES is subject to microwave and dc electric field simultaneously.
The electron state transition can be viewed as either a jump in energy due to MW absorption or a jump in space due to elastic scattering by impurities under dc excitation ~\cite{ZudovPRL2007}.
From this point of view, HIRO and MIRO are considered as separate processes .

Most experiments about MIRO and ZRS are focused on high-frequency ($f > 30$ GHz) MW irradiation with high energy photon.
The limitation is that MW signals cannot be transmitted below the cut-off frequency within the rectangular waveguide.
In this paper we expand the MW frequency range of MIRO by using a linear dipole antenna, and focus on the magnetoresistance oscillations under low-frequency ($f < 20$ GHz) MW irradiation.
In contrast to previous results ~\cite{Willet2004}, we observe abundant multi-photon processes of MIRO with decreasing MW frequency and a new Shubnikov de-Haas (SdH)-like oscillation under ultra-low frequency ($f < 4$ GHz).
We interpret the new oscillation by the theory of ac-HIRO or the multi-photon processes of MIRO, whereby the two different non-equilibrium microscopic mechanisms could be unified under certain conditions.
The low frequency results provide good supplements to previous MIRO experiments.

Our experiments are carried out in a top-loading He3 refrigerator with a base temperature of 0.3 K.
The wafer with a high-quality ($\mu \sim 3\times 10^{7}$ cm$^{2}$/Vsec) GaAs/Al$_{0.24}$Ga$_{0.76}$As quantum well (QW) is grown by molecular-beam epitaxy.
The 28 nm wide QW is located about 320 nm beneath the sample surface.
The Hall bar sample is defined by UV-lithography and wet etching.
Ohmic contacts are made by a 43/ 30/ 87 nm stack of Ge/ Pd/ Au metals with an annealing process at 450 $^{\circ}$C.
The microwave signal is generated by a continuous wave generator Anritsu MG3690C, and guided to the base via a semi-rigid coaxial cable, radiating the sample by a linear dipole antenna hung over the sample.
The microwave frequency ranges from 8 MHz to 70 GHz.
The resistance is measured by applying a low-frequency (17 Hz) external current $I = 100$ nA through the Hall bar, and probing the voltage drop of two contacts.

\begin{figure}
\includegraphics[width=0.8\linewidth]{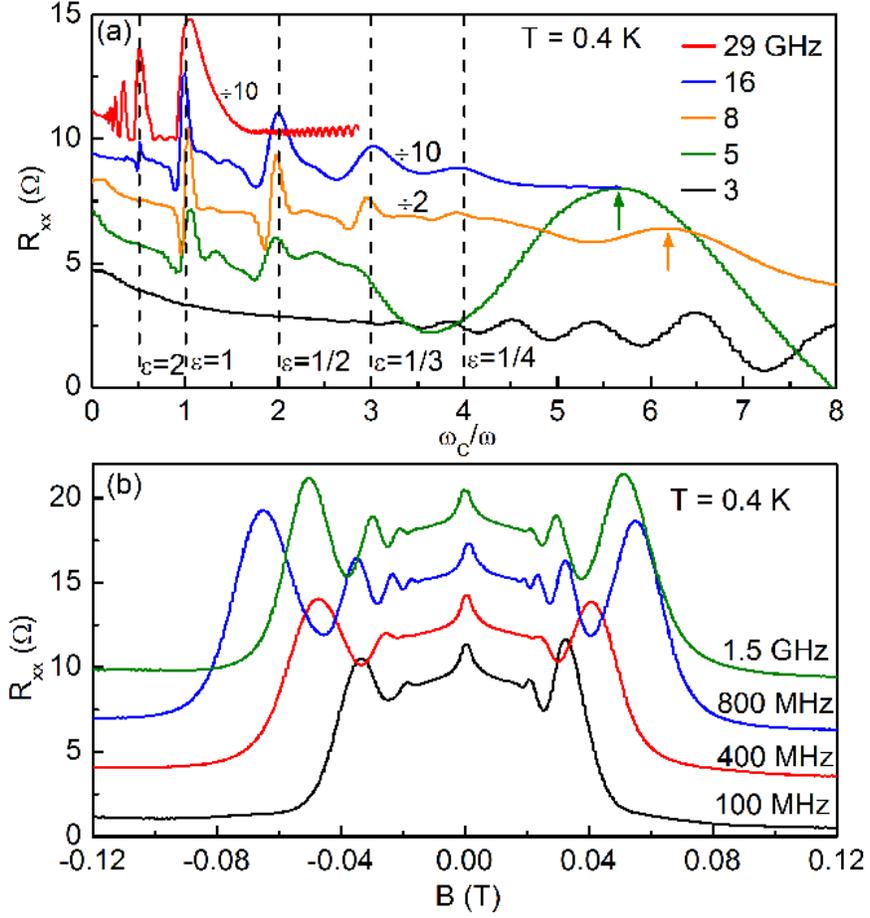}
\caption{(Color online). (a) Magnetoresistances under MW irradiation from high frequency (29 GHz) to low frequency (3 GHz) are plotted against $\omega_{c}/ \omega$, and the electron effective mass $m^{*}$ is taken to be $0.067 m_{e}$. The dashed lines indicate the multi-photon and harmonic process. (b) $R_{xx}$ under low frequency MW irradiation ($< 1.5$ GHz). All the traces with vertically offset are shown for clarity.}
  \label{FIG1}
 \end{figure}

Fig. 1(a) illustrates the results of magnetoresistance under MW irradiation from 3 GHz to 29 GHz.
At $f = 29$ GHz, MIRO and ZRS are clearly observed.
The major peaks are located near $\varepsilon = \omega / \omega_{c} = 1, 2, 3$, ..., which indicates that the harmonic process is dominant --- an electron absorbs one MW photon and jumps one or multiple Landau levels.
However, with a lower frequency $f = 16$ GHz, the resistance peaks near  $\varepsilon > 2$ disappear, but peaks at  $\varepsilon = 1/2, 1/3, 1/4$ develop, these are called multi-photon processes ~\cite{Zudov-Du2006}, in which an electron absorbs two or more MW photons and jumps one Landau level.
Multi-photon processes only exist at low MW frequency ($f < 30$ GHz).
At $f = 8$ GHz, the peak near  $\varepsilon = 2$ disappears, and the amplitude of MIRO reduces rapidly.
Meanwhile, a small resistance peak near  $\varepsilon = 1/6.2$ arises (marked by the orange color arrow), which cannot be explained by the regime of MIRO.
The new peak (marked by the green color arrow) expands at $f = 5$ GHz, and grows stronger than MIRO.
Finally, MIRO disappears completely at $f = 3$ GHz, and the new oscillation becomes dominant.
In general, with decreasing MW frequency and photon energy, MIRO decays gradually while a new oscillation emerges and develops.
Fig. 1 (b) shows $R_{xx}$ vs. $B$ under low MW frequency irradiation with $f = 1.5$ GHz, 800 MHz, 400 MHz, 100 MHz, respectively.
The new oscillations persist in a wide range of MW irradiation frequency, but do not depend on MW frequency.
Like SdH oscillations, they are roughly $1/B$ periodic.
And similar to MIRO and HIRO, the new resistance oscillations are symmetric with positive and negative magnetic fields, appearing only at very high Landau levels ($B < 0.01$ T).

\begin{figure}
\includegraphics[width=0.8\linewidth]{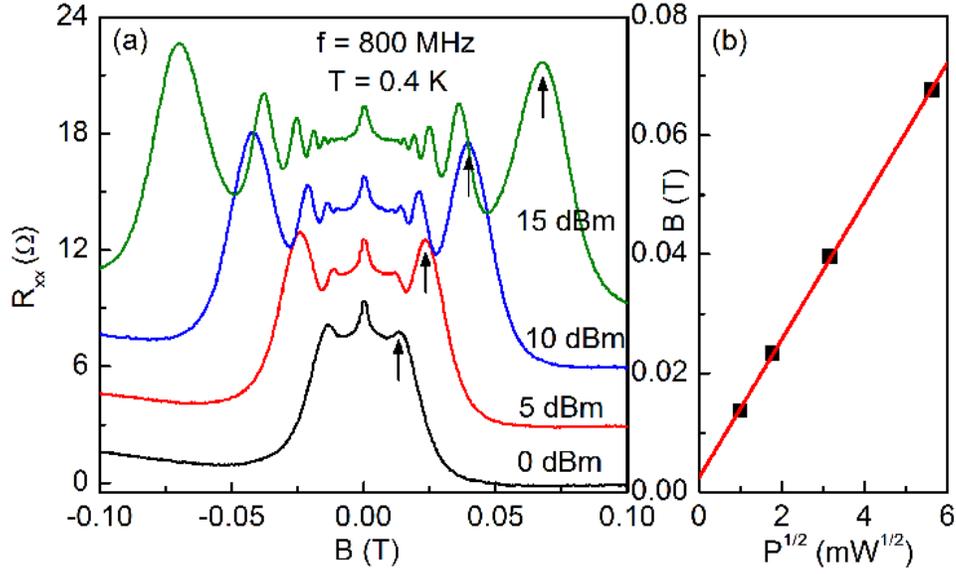}
\caption{(Color online). (a) Power-dependent $R_{xx}$ under 800 MHz microwave irradiation (with vertical offset for clarity). The upward arrows mark the largest resistance peak of the oscillations. (b) The $B$ positions of the marked peaks versus the square root of MW power (black squares). The red line is the linear fitting curve.}
  \label{FIG1}
 \end{figure}

The MW power-dependent results of the new oscillations for $f = 800$ MHz are shown in Fig. 2(a).
The MW power values are obtained from the wave generator; the power loss in coaxial cables and the efficiency of the linear dipole antenna are not taken into account.
The new oscillations are strongly dependent on MW power --- at low power, the oscillations are weak and appears only at low magnetic fields.
When the power increases, the oscillations extends to higher magnetic fields, and the number of peaks grows as well as the amplitudes of the oscillations.
The $B$-field positions of the highest resistance peak (marked by upward arrows) versus the square root of MW power are shown in Fig. 2(b).
The linear relationship indicates that the $B$-field position of the peak is proportional to the electric component of microwave (or electromagnetic (EM)) field.
The power-dependent features are different from MIRO, in which only the amplitudes change with MW power but the $B$-field positions remain constant.
For HIRO, the oscillation is dependent on the dc-bias current density, akin to the new oscillations that are $EM$-field dependent.
All these features suggest that the new oscillations may pertain to HIRO.

\begin{figure}
\includegraphics[width=0.8\linewidth]{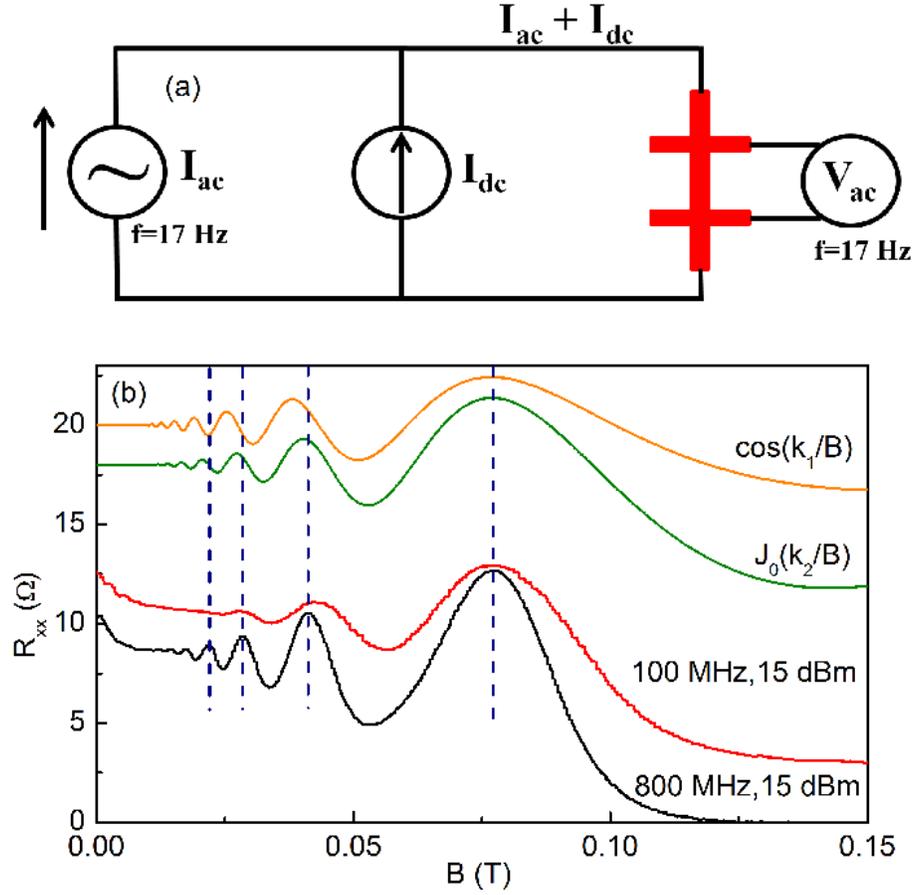}
\caption{(Color online). Panel (a) The diagram for the electrical measurement of HIRO. (b) The comparison of $R_{xx}$ with Cosine function and Bessel function. All traces are horizontally scaled to align the largest oscillation peaks; and vertical offset are provided  for clarity. The dashed lines indicate the peaks of the 800 MHz trace.}
  \label{FIG1}
 \end{figure}

Fig. 3(a) displays a circuit diagram for the electrical measurement of HIRO.
A constant dc current $I_{dc}$ is applied through the Hall bar sample, along with a low frequency (17 Hz) modulation current $I_{ac}$ of 100 nA.
The voltage drop between the two contacts is recorded by a lock-in amplifier at the modulation frequency.
The result is called the differential magnetoresistance at a given dc-bias: $r_{xx} = V_{ac} / I_{ac} = (\partial V/ \partial I)_{dc}$ .
When the 2DES sample is irradiated by MW, an alternating current is excited in the Hall bar by the $EM$-fields.
The excited current plays a role of the bias current, causing ac-HIRO effect.

For the sake of simplification, the magnetoresistance of dc-bias-induced HIRO can be described with a Cosine function:  $\Delta R = A\delta^{2} \cos (\eta J_{dc}/B)$, where $A$ is the amplitude, $\delta = \exp(-\pi /\omega_{c}\tau_{q})$ is the Dingle factor from the SdH oscillations,  $\tau_{q}$ is the quantum lifetime of electrons, and $J_{dc}$ is the dc-bias current density.
The factor can be expressed as $\eta = 4 \pi m^{*}\sqrt{\frac{2 \pi}{n}} /e^{2}$, in which $n$ is the two-dimensional electron density.
If the bias is an alternating current ~\cite{BykovPRB2005}, $J_{dc}=j_{0}\cos(\omega t)$, where $j_{0}$ and $\omega$ are the amplitude and the frequency of the alternating current respectively.
Thus $\Delta R$ changes with time rapidly.
But in our measurement, the lock-in amplifier is locked at a low frequency of 17 Hz, and will only reach the zero frequency harmonic $\Delta R_{0}$ from $\Delta R$, because only $\Delta R_{0}$ provides a term oscillating at 17 Hz.
$\Delta R_{0}$ is derived from Fourier transform:
\begin{equation}
 \label{1}
    \Delta R_{0} = \frac{1}{T} \int_{-T/2}^{T/2} \Delta R dt = A \delta^{2}J_{0}(\eta \frac{j_{0}}{B}).
  \end{equation}
Therefore, the ac-HIRO result can be described as a simple Bessel function $J_{0}(x)$. The oscillation depends on the amplitude of the alternating current and the magnetic field, and $j_{0}$ is determined by the intensity of $EM$-field or MW power, in accord with the power-dependent results in Fig. 2.

Fig. 3(b) shows the comparison of $R_{xx}$ under MW irradiation with Cosine function and Bessel function.
The traces of 100 MHz and 800 MHz irradiation are both in line with the Bessel function trace, as opposed to the Cosine function trace.
The model of ac-HIRO supports the explanation of the new SdH-like oscillation.

\begin{figure}
\includegraphics[width=0.8\linewidth]{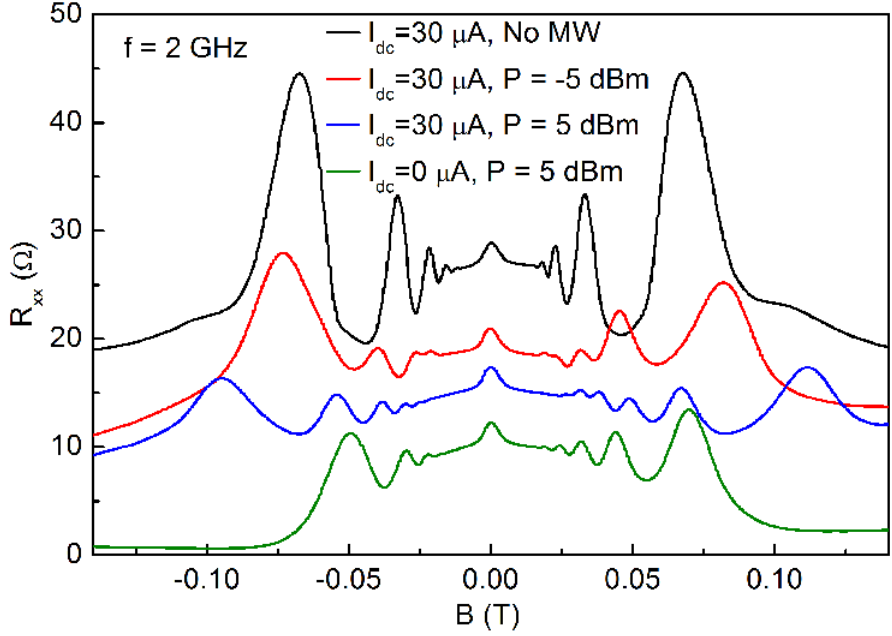}
\caption{(Color online). Magnetoresistance under MW irradiation for $f = 2$ GHz, with a dc-bias current passing through the Hall bar sample. The two kinds of oscillations merge into a new type of oscillation.}
  \label{FIG1}
 \end{figure}

To further demonstrate the close relationship between the new oscillation and HIRO, we study the magnetoresistance of the sample subject to microwave irradiation and a dc-bias current simultaneously.
The results are shown in Fig. 4.
Without MW irradiation, a strong dc-HIRO arises, which can be described as Cosine function.
Meanwhile, with increasing MW power, the oscillation expands to higher magnetic field and the amplitude decreases.
We find that the MW induced new oscillation and dc-induced HIRO merge into a combined oscillation, which suggests that the mechanism of the new oscillation is consistent with HIRO.
When the sample is driven by low frequency MW, the bias can be expressed as a function of driven frequency: $J_{dc}=j_{1}+j_{0}\cos(\omega t)$, where $j_{1}$ and $j_{0}$ are the amplitudes of the dc-bias current and the excited alternating current respectively.
And the zero frequency harmonic is derived as: $\Delta R_{0} = A \delta ^{2} J_{0}(\eta \frac{j_{0}}{B}) \cos (\eta \frac{j_{1}}{B})$.

The combined oscillation is determined by both $j_{1}$ and $j_{0}$.
If the dc-bias current density $j_{1}$ is fixed, the MW irradiation can be viewed as a Bessel modulation on HIRO.
Consequently, the $B$-field coordinate of the oscillation peaks increases with the MW power.
It is strong evidence that the new oscillation originates from HIRO.

\begin{figure}
\includegraphics[width=0.8\linewidth]{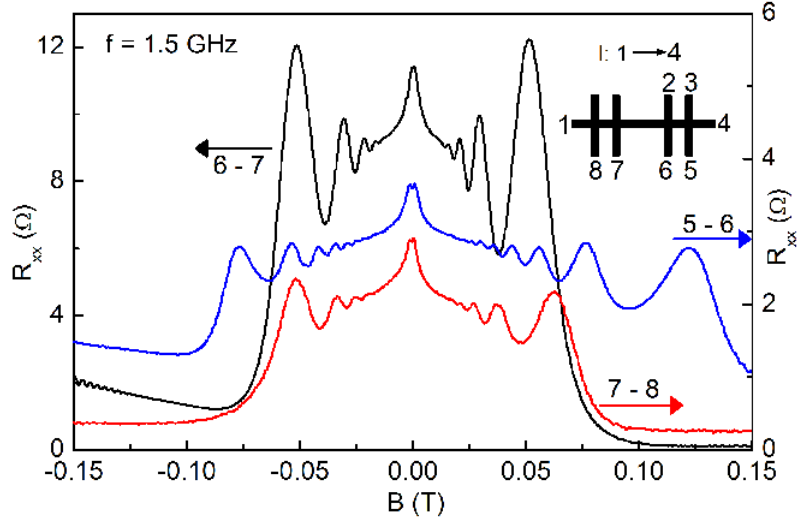}
\caption{(Color online). Magnetoresistance results of different contact configurations under MW irradiation. The inset shows the diagram of the Hall bar sample, and the excitation current $I_{ac}$ is applied from contact 1 to 4.}
  \label{FIG1}
 \end{figure}

The hitherto magnetoresistance results in this report are measured from the contacts 6 and 7 of the Hall bar sample (inset in Fig. 5).
Fig. 5 presents $R_{xx}$ results from different contact configurations under MW irradiation. We find that the oscillations are strongly dependent on contact configurations, different from MIRO and HIRO.
Under MW irradiation, every Hall bar arm can be considered as an antenna, and an alternating bias current is excited inside.
Since the distance between the dipole antenna and the sample is less than 20 mm, and the size of the sample is larger than 1 mm, the $EM$-field near the Hall bar cannot be described as plane wave from far-field approximation, hence the amplitude and the phase of the $EM$-field change with spatial positions.
For instance, the equivalent alternating bias currents of contact configurations 7-8, 6-7, and 5-6 are not equal.
It gives rise to the contact-dependent oscillations.

The MW signal in the ac-HIRO mechanism is considered as excited bias current. We can also explain the new oscillations in the view of microwave photons.
Among the various theoretical models proposed for MIRO and ZRS, Shi and Xie came up with a quantum tunneling junction model ~\cite{Shi-Xie2003}, by which a periodic voltage $V_{ac}$ (or electric field) can be induced by microwave (or rf) $EM$-field.
The obtained photon-assisted transport result of conductivity is derived as:
\begin{equation}
 \label{2}
    \frac{\sigma}{\sigma_{0}} = \sum_{n = -\infty}^\infty J_{n}^{2}(\frac{\Delta}{\hbar \omega}) [1 + \frac{\lambda^{2}}{2} \cos(2 \pi n\frac{\omega}{\omega_{c}}) -n \pi \lambda^{2}\frac{\omega}{\omega_{c}} \sin (2 \pi n \frac{\omega}{\omega_{c}})].
  \end{equation}
$\sigma$ and $\sigma_{0}$ are the conductance of the system with and without MW irradiation.
The conductivity can be expressed as a function of MW power and frequency: $\Delta$ is the $EM$-field intensity, $\omega$ is the MW frequency, $\lambda = 2 \exp (- \pi \omega_{c} \tau_{q})$ is the Dingle factor, and $J_{n}(x)$ is the Bessel function of $n$-th order.
The numerical results show that with increasing MW power, the minimum of the conductance oscillation becomes negative and zero resistance state emerges.
When the power increases further, multi-photon process sets in. Equation (2) can also be simplified as:
$\frac{\sigma}{\sigma_{0}} = 1 + \frac{\lambda^{2}}{2} J_{0}(\frac{2\Delta}{\hbar \omega} \sin(\frac{\pi \omega}{\omega_{c}})) - \frac{\pi \lambda^{2} \Delta}{\hbar \omega_{c}} \cos(\frac{\pi \omega}{\omega_{c}}) J_{1}(\frac{2\Delta}{\hbar \omega} \sin(\frac{\pi \omega}{\omega_{c}}))$.

At the low MW frequency limit ($\omega / \omega_{c} \rightarrow 0$), with $\sin (\frac{\pi \omega}{\omega_{c}}) \approx \frac{\pi \omega}{\omega_{c}}$, $\cos(\frac{\pi \omega}{\omega_{c}}) \approx 1$, the change of conductance is:
\begin{equation}
 \label{3}
    \frac{\Delta \sigma}{\sigma_{0}} = \frac{\lambda^{2}}{2} J_{0}(\frac{2\pi \Delta}{\hbar \omega_{c}}) - \frac{\pi \lambda^{2} \Delta}{\hbar \omega_{c}} J_{1}(\frac{2\pi \Delta}{\hbar \omega_{c}}) \approx \frac{\lambda^{2}}{2} J_{0}(\frac{2\pi \Delta}{\hbar \omega_{c}}) = \frac{\lambda^{2}}{2} J_{0}(\frac{2\pi m^{*}}{e \hbar} \frac{\Delta}{B}).
  \end{equation}
The result indicates that the conductance is not dependent on MW frequency, but depends on $EM$-field intensity at low frequency limit.
The quantum tunneling junction model provides an explanation: under low MW frequency irradiation with low photon energy, an electron absorbs multiple photons and jumps one Landau level.
The new oscillation can be regarded as a multi-photon process of MIRO at low MW frequency limit.
Equation (3) on low-frequency limit of MIRO has the same form as Eq. (1) on HIRO phenomenon, where $\lambda = 2 \delta$, and $\Delta$ corresponds to $j_{0}$.
We explain the new oscillations with either multi-photon limit of MIRO or ac-HIRO.
MIRO describes electron transitions from the photon absorption in energy, while HIRO is the transitions from the elastic scattering in space.
Low frequency MW is in between the two limits of bias for MIRO (high-frequency MW) and HIRO (dc-bias).
These two mechanisms can be integrated into one for low frequency MW irradiation.

In summary, in our ultrahigh mobility n-type GaAs/AlGaAs quantum well, we expand the MW frequency range of MIRO and focus on magnetoresistance oscillations under low-frequency ($< 20$ GHz) MW irradiation.
We find that with decreasing frequency, multi-photon assisted MIRO is observed, and when MIRO disappears gradually a new SdH-like oscillation develops.
The new oscillation is dependent on MW power rather than MW frequency.
We provide a qualitative analysis of the observation in terms of alternating bias current induced HIRO and multi-photon process of MIRO at low MW frequency limit, whereby the two different non-equilibrium mechanisms can be unified under certain conditions.
It is a desirable approach to perceive HIRO and MIRO in a unified theory.
Our findings contribute to the understanding of the relationship between MIRO and HIRO.


\begin{thebibliography}{text}
\bibitem{Zudov2001} M. A. Zudov, R. R. Du, J. A. Simmons, and J. L. Reno, {\it Phys. Rev. B} {\bf 64}, 201311(R) (2001).
\bibitem{YeAPL2001} P. D. Ye, L. W. Engel, D. C. Tsui, J. A. Simmons, J. R. Wendt, G. A. Vawter, and J. L. Reno, {\it Appl. Phys. Lett.} {\bf 79}, 2193 (2001).
\bibitem{Mani} R. G. Mani, J\"{u}rgen H. Smet, Klaus von Klitzing, Venkatesh Narayanamurti, William B. Johnson, and Vladimir Umansky, {\it Nature} {\bf 420}, 646 (2002).
\bibitem{Zudov} M. A. Zudov, R. R. Du, L. N. Pfeiffer, and K. W. West, {\it Phys. Rev. Lett.} {\bf 90}, 046807 (2003).
\bibitem{YCL2002} C. L. Yang, J. Zhang, R. R. Du, J. A. Simmons, and J. L. Reno, {\it Phys. Rev. Lett.} {\bf 89}, 076801 (2002).
\bibitem{ZudovPRB2007} W. Zhang, H.-S. Chiang, M. A. Zudov, L. N. Pfeiffer, and K. W. West, {\it Phys. Rev. B} {\bf 75}, 041304 (2007).
\bibitem{Durst2003} Adam C. Durst, Subir Sachdev, N. Read, and S. M. Girvin, {\it Phys. Rev. Lett.} {\bf 91}, 086803 (2003).
\bibitem{Dimitriev2005} I. A. Dmitriev, M. G. Vavilov, I. L. Aleiner, A. D. Mirlin, and D. G. Polyakov, {\it Phys. Rev. B} {\bf 71}, 115316 (2005).
\bibitem{HatkePRB2008} A. T. Hatke, H.-S. Chiang, M. A. Zudov, L. N. Pfeiffer, and K. W. West, {\it Phys. Rev. B} {\bf 77}, 201304(R) (2008).
\bibitem{ZudovPRL2007} W. Zhang, M. A. Zudov, L. N. Pfeiffer, and K. W. West, {\it Phys. Rev. Lett.} {\bf 98}, 106804 (2007).
\bibitem{HatkePRL2008} A. T. Hatke, H.-S. Chiang, M. A. Zudov, L. N. Pfeiffer, and K. W. West, {\it Phys. Rev. Lett.} {\bf 101}, 246811 (2008).
\bibitem{Willet2004} R. L. Willett, L. N. Pfeiffer, and K.W. West, {\it Phys. Rev. Lett.} {\bf 93}, 026804 (2004).
\bibitem{Zudov-Du2006} M. A. Zudov, R. R. Du, L. N. Pfeiffer, and K. W. West, {\it Phys. Rev. B} {\bf 73}, 041303(R) (2006).
\bibitem{BykovPRB2005} A. A. Bykov, J. Q. Zhang, Sergey Vitkalov, A. K. Kalagin, and A. K. Bakarov, {\it Phys. Rev. B} {\bf 72}, 245307 (2005).
\bibitem{Shi-Xie2003} Junren Shi and X. C. Xie, {\it Phys. Rev. Lett.} {\bf 91}, 086801 (2003).


\end{thebibliography}
\end{document}